# Rosen-Zener model in cold molecule formation

A. Ishkhanyan[1], R. Sokhoyan[1], B. Joulakian[2], and K.-A. Suominen[3]

[1]*Institute for Physical Research of NAS of Armenia, 0203 Ashtarak-2, Armenia*
[2]*Institut de Physique, Université Paul Verlaine - Metz, LPMC, 57078 Metz Cedex 3, France*
[3]*Department of Physics, University of Turku, FI-20014 Turun yliopisto, Finland*

The Rosen-Zener model for association of atoms in a Bose-Einstein condensate is studied. Using a nonlinear Volterra integral equation, we obtain an analytic formula for final probability of the transition to the molecular state for weak interaction limit. Considering the strong coupling limit of high field intensities, we show that the system reveals two different time-evolution pictures depending on the detuning of the frequency of the associating field. For both limit cases we derive highly accurate formulas for the molecular state probability valid for the whole range of variation of time. Using these formulas, we show that at large detuning regime the molecule formation process occurs almost non-oscillatory in time and a Rosen-Zener pulse is not able to associate more than one third of atoms at any time point. The system returns to its initial all-atomic state at the end of the process and the maximal transition probability $1/6$ is achieved when the field-intensity reaches its peak. In contrast, at small detuning the evolution of the system displays large-amplitude oscillations between atomic and molecular populations. We find that the shape of the oscillations in the first approximation is defined by the field detuning only. Finally, a hidden singularity of the Rosen-Zener model due to the specific time-variation of the field amplitude at the beginning of the interaction is indicated. It is this singularity that stands for many of the qualitative and quantitative properties of the model. The singularity may be viewed as an effective resonance-touching.

## 1. Introduction

Recent exciting achievements in ultralow-temperature atomic and molecular physics (see, e.g., [1-3]) are due to intensive developments and artful applications of several experimental techniques. Among those, the optical photoassociation [4] and the magnetic Feshbach resonance [5] currently became two powerful tools for molecule formation from ultracold atoms [2-5]. Much flexibility emerges here from combination of these processes with level-crossings adjusted by time-variation of the applied electromagnetic field's phase and amplitude [2-5]. This point suggests strong motivation for systematic exploration of pulse shape and detuning modulation effects in nonlinear quantum systems such as the Bose-Einstein condensate [1] and degenerate Fermi gases [3]. However, while for linear quantum systems several models have been developed to cover various possible situations, most of the theory for the nonlinear case is so far focused on the Rabi non-crossing [6] and Landau-Zener linear-crossing [7] models (see, e.g., [8-12]). Since consideration of more realistic models, as already known from the linear theory [13-14], can add much both to the understanding of basic physical processes and to the development of efficient experimental tools for precise control of cold atom motion, there is a pronounced need to explore models other than the mentioned two.

For the non-crossing models the next after the basic constant-amplitude Rabi one comes the Rosen-Zener model [15] of finite pulse duration, when the detuning is supposed constant while the field amplitude varies in time according to the hyperbolic secant low. Though in the limits of the model considered here the cold molecule formation processes via photoassociation or a Feshbach resonance are mathematically treated in equivalent manner, this field configuration is directly relevant to the photoassociation only. This is because in the case of a magnetic resonance the coupling term (i.e., the pulse duration, if optical terminology is used) can not be adjusted - it corresponds to some given hyperfine coupling. On contrary, in photoassociation the pulse duration can not be infinite (this would mean infinite energy). Hence, finite pulse duration should necessarily be discussed if experimental realization is assumed. To this end, the accumulated knowledge from the linear theory suggests that one should be careful with the optical pulse inclusion and shutdown scenarios – the particular form of the time-variation of the field amplitude plays a substantial role. A well discussed shape of such a time-variable pulse in the linear theory is the Rosen-Zener hyperbolic-secant model. This is a motivation for exploring the Rosen-Zener field-configuration for the photoassociation.

One should note, however, that this model is applied, though indirectly, to the Feshbach resonance as well. This is achieved by applying a transformation of the independent variable (time) that changes the governing equations to a constant-amplitude form (see below). Changing to the constant-amplitude form turns the model into a variable-detuning field-configuration. Yet, strictly speaking, the model remains non-crossing. In the meantime, this constant-amplitude form reveals a prominent property of the model, namely, a hidden singularity due to the speed of the field inclusion at $t = -\infty$. It is this singularity that makes a major difference of this model from the Rabi one, which does not reveal the different evolution scenarios inherent for the Rosen-Zener model as discussed below. The mentioned singularity effectively acts as a resonance-touching. Finally, it should be noted that the constant-amplitude form of the model makes it relevant to several recent experiments. Thus, the model is equally useful for the magneto-association via Feshbach resonances.

In the present paper we explore both the weak and strong coupling regimes for the Rosen-Zener field-configuration comparing the results with those for the linear Rosen-Zener model [15] and the nonlinear Rabi problem [8].

In the quasi-resonance approximation, the semiclassical equations describing two-mode one-color photo- or magneto-association of an atomic Bose-Einstein condensate have the form of the following system of coupled nonlinear equations for the probability amplitudes of the atomic and molecular states $a_1$ and $a_2$ [16-18]:

$$i\frac{da_1}{dt} = U(t)e^{-i\delta(t)}a_2\bar{a}_1, \quad i\frac{da_2}{dt} = \frac{U(t)}{2}e^{i\delta(t)}a_1 a_1, \tag{1}$$

where $t$ is the time, $U(t)$ is the Rabi frequency, and $\delta(t)$ is the detuning modulation function. These equations are often faced in different field theories with a Hamiltonian containing a term of the form $a_2^+ a_1 a_1$, for instance, in controlling the interaction strength by a Feshbach resonance [5,18], in second harmonic generation in nonlinear optics [19], etc. System (1) preserves the total number of particles that we normalize to unity: $|a_1|^2 + 2|a_2|^2 = \text{const} = 1$. We will consider a condensate initially being in pure atomic state: $a_1(-\infty)=1$, $a_2(-\infty)=0$. All the parameters involved in (1) are supposed dimensionless.

We study the following field configuration known as the Rosen-Zener model [15]:

$$U(t) = U_0 \text{sech}(t), \quad \delta(t) = 2\delta_0 t. \tag{2}$$

In the analysis below the following linear analog of system (1) is used:

$$i\frac{da_{1L}}{dt} = U(t)e^{-i\delta(t)}a_{2L}, \quad i\frac{da_{2L}}{dt} = U(t)e^{+i\delta(t)}a_{1L}, \tag{3}$$

with the same functions $U(t), \delta(t)$, initial condition $a_{2L}(-\infty)=0$ and motion integral of the form $|a_{1L}|^2 + |a_{2L}|^2 = \text{const} = I_L$. Here, to ensure coincidence of the solutions to systems (1) and (3) in the vicinity of $t=-\infty$, it should be taken $I_L = 1/4$, which leads to $|a_{1L}(-\infty)| = 1/2$. Note that the solution of system (3) for $I_L = 1$ is written as

$$\begin{aligned}
a_{1RZ} &= {}_2F_1(U_0, -U_0; 1/2 + i\delta_0; x), \\
a_{2RZ} &= -iU_0((1-x)/x)^{-i\delta_0}\sqrt{x(1-x)} \cdot {}_2F_1(1+U_0, 1-U_0; 3/2+i\delta_0; x)/(1/2+i\delta_0),
\end{aligned} \tag{4}$$

where $x = (1+\tanh(t))/2$, and ${}_2F_1(\alpha, \beta; \gamma; x)$ is the Gauss hypergeometric function [20]. Hence, accurate to a phase factor, the solution for system (3) satisfying the normalization $I_L = 1/4$ is $a_{1L} = a_{1RZ}/2$, $a_{2L} = a_{2RZ}/2$.

The final (at $t \to +\infty$) probability of transition to the second level is given by the following nice formula by Rosen and Zener [15]:

$$P_{RZ} = [\sin(\pi U_0)]^2 \cdot [\text{sech}(\pi\delta_0)]^2. \tag{5}$$

This formula states the well-known $\pi$-theorem [13] according to which the system returns to the initial state $(a_1, a_2) = (1,0)$ if $U_0 = n$ with $n = 0,1,2...$, and reaches the highest transition probability possible for the given fixed detuning at $U_0 = 1/2 + n$ ($P_{RZ}^{max} = [\text{sech}(\pi \delta_0)]^2$). Note that the system is completely inverted at exact resonance only.

The numerical solutions to nonlinear and linear systems are compared in Fig.1. As it is seen, the nonlinear behavior displays considerable deviations from the linear case. First, at exact resonance the dependence of the final transition probability on the Rabi frequency in the nonlinear case is monotonic. Second, while at non-zero detuning atom/molecule oscillations are always observed as the field amplitude is increased, the $\pi$-theorem is no longer valid. However, importantly, at fixed detuning the final transition probability depends nearly periodically on the field amplitude and approximately periodic *returns* to the initial state are observed. (Therefore, it is likely that a changed form of the $\pi$-theorem holds in this nonlinear case as well.) This is demonstrated in Fig.2. Furthermore, examining the graphs in this figure, we see that the oscillation shape, amplitude and frequency are changed depending on the detuning. Clearly, the oscillation nature is close to that of the nonlinear Rabi-solution (see, e.g., [8]). Finally, we note that in the nonlinear case the transition probability decreases considerably faster as the detuning is increased, becoming negligible already at $\delta_0 \approx 1$.

Our study is based on two different *exact* nonlinear equations written for the molecular state probability $p = |a_2|^2$, a Volterra integral equation and a third-order differential equation. Being equivalent in general, these two equations are effective if applied to opposite limits: the first equation is productive at weak interaction while the second one works at strong coupling.

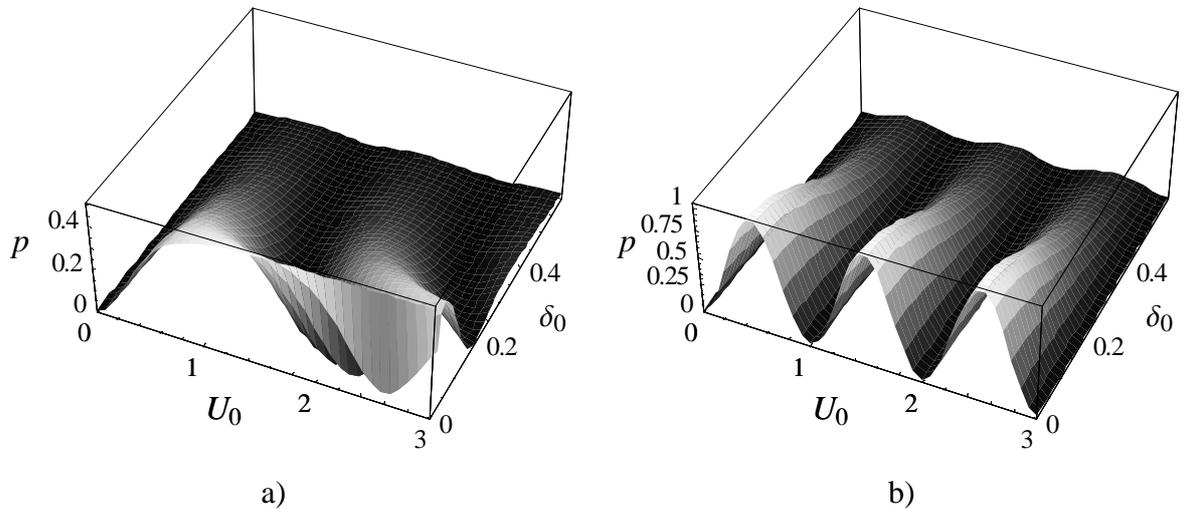

a)                                                                 b)
Fig.1. Final transition probability: a) nonlinear problem and b) linear problem.

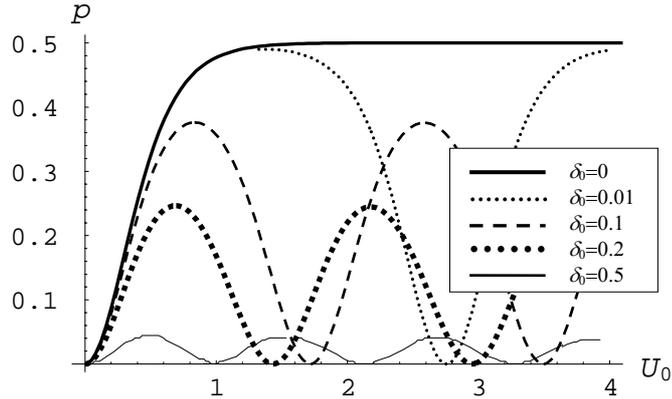

Fig.2. Nonlinear Rosen-Zener model. At fixed non-zero detuning the transition probability depends almost periodically on the field amplitude. The oscillation shape, amplitude and frequency undergo changes analogous to those observed for the nonlinear Rabi-problem.

We start from the weak coupling regime of small field intensities that is often encountered situation under the current experimental conditions. Using the nonlinear Volterra integral equation, we show that an accurate approximate solution for this limit can be constructed, using Picard's successive approximations, in terms of the solution to the linear quantum-optical problem. We determine the final conversion probability and show that because of the inherent properties of the Rosen-Zener model under consideration the strict limit of weak nonlinearity (when no essential deviations from the linear evolution are observed) corresponds to smaller field intensities as compared with the Landau-Zener case. We discuss the specific reasons for such behavior and construct an approximation that is valid also for the intermediate regime of moderate coupling strength.

Further, we pass to the strong coupling limit of high field intensities and show that the system reveals two different time-evolution pictures depending on the frequency detuning of the associating field. At large detuning the molecule formation process occurs almost non-oscillatory in time. In contrast, at small detuning the evolution of the system displays strongly pronounced large-amplitude Rabi-type oscillations. The third-order differential equation in each case is reduced to a limit equation of lower order. In the case of large detuning this equation is of the first order, while in the small detuning case it is an effective Rabi-equation of the second order. Using these limit equations, we derive two accurate approximate formulas for the molecular state probability applicable to the two mentioned regimes. The results show that in the large detuning regime the system always returns to the initial all-atomic state independently on the field intensity, hence, the final molecule formation efficiency in this case is nearly zero. In the small detuning regime, because of large-amplitude oscillations, the Rabi frequency (or, equivalently, the Rosen-Zener pulse area) should be adjusted in order to achieve efficient conversion.

## 2. Weak coupling limit

Consider the transformation of the independent variable $dz/dt = \text{sech}(t)$ that changes system (1) to the following ***constant-amplitude*** form

$$i\frac{da_1(z)}{dz} = U_0 e^{-i\delta(z)} \bar{a}_1(z) \cdot a_2(z), \quad i\frac{da_2(z)}{dz} = \frac{U_0}{2} e^{+i\delta(z)} a_1(z) \cdot a_1(z), \tag{6}$$

where $z = \pi/2 + 2\arctan(\tanh(t/2))$ ($z \in [0,\pi]$) and

$$\delta(z) = 4\delta_0 \, \text{arctanh}\left(\tan\left(\frac{z}{2} - \frac{\pi}{4}\right)\right) \Rightarrow \delta_z(z) = \frac{2\delta_0}{\sin(z)}. \tag{7}$$

To treat the weak coupling limit of such problems with arbitrary detuning and constant Rabi frequency $U_0$, we have earlier developed an appropriate mathematical approach based on the reduction of system (6) to the following nonlinear Volterra integral equation [21] for the molecular state probability $p(z) = |a_2(z)|^2$ [12]:

$$p(z) = \frac{\lambda}{4} f(z) - 4\lambda \int_0^z K(z,\eta)\left(p(\eta) - \frac{3}{2}p^2(\eta)\right) d\eta, \tag{8}$$

where $\lambda = U_0^2$ and the kernel, $K(z,\eta)$, and the forcing function, $f(z)$, are given as

$$K(z,\eta) = (C_\delta(z) - C_\delta(\eta))\cos(\delta(\eta)) + (S_\delta(z) - S_\delta(\eta))\sin(\delta(\eta)), \tag{9}$$

$$f(z) = C_\delta^2(z) + S_\delta^2(z), \quad C_\delta(z) = \int_0^z \cos(\delta(\xi))d\xi, \quad S_\delta(z) = \int_0^z \sin(\delta(\xi))d\xi. \tag{10}$$

Note that if the term proportional to $p^2$ is omitted, Eq. (8) turns into an exact equation equivalent to the linear system (3). In the case of ***weak coupling*** ($U_0^2 < 1$), a series solution to the problem is constructed by means of Picard's successive approximations [21] to equation (8). Furthermore, noting that the first three terms of this expansion and that of corresponding linear integral equation coincide, it is possible to construct a faster converging series using the substitution $p = p_L + u$ where $p_L = |a_{2L}|^2$. For the function $u(z)$ we get a new integral equation of the Hammerstein type [21]:

$$u(z) = 6\lambda \int_0^z K(z,\eta) p_L^2 d\eta - 4\lambda \int_0^z K(z,\eta)\left[(1 - 3p_L)u - \frac{3}{2}u^2\right] d\eta. \tag{11}$$

It is not difficult to see that it is sufficient to take only the first term of Eq. (11). Thereby, the approximate solution to system (6) is written by means of the solution to linear system (3):

$$p(z) = p_L(z) + 6\lambda \int_0^z K(z,\eta) p_L^2 d\eta. \qquad (12)$$

This formula is checked to be rather accurate in an appropriate range of variation of $\lambda = U_0^2$.

Now consider how to calculate the integral in Eq. (12). Note that to achieve a preset accuracy in powers of $\lambda$, the approximation of $p_L$ by a finite number of terms of its Picard's series can be used. Restricting to the accuracy up to $O(\lambda^4)$ (the first order of the expansion), one may put $p_L(z) \approx \lambda f(z)/4$. To improve this approximation, a correction factor can be introduced, thereby applying an approximation of the form $p_L(z) \cong A f(z)$. Furthermore, the functions $C_\delta$ and $S_\delta$ are explicitly determined by considering an auxiliary integral:

$$F(z) \equiv C_\delta + iS_\delta = \int e^{i\delta(\xi)} d\xi = B_y(1/2 + i\delta_0, 1/2 - i\delta_0), \quad y = \sin^2(z/2). \qquad (13)$$

Hence

$$C_\delta = \text{Re}[B_y(1/2 + i\delta_0, 1/2 - i\delta_0)], \quad S_\delta = \text{Im}[B_y(1/2 + i\delta_0, 1/2 - i\delta_0)]. \qquad (14)$$

These functions are shown in Fig.3.

Then, the correction term $u$ at $z = \pi$ ($t = +\infty$) is readily calculated. The result reads $u(\pi/2) \approx 6\lambda A^2 C_\infty^3$, where $C_\infty = \pi \text{sech}(\pi\delta_0)$. Hence, for $A$ chosen as $A = \lambda/4$, the solution to nonlinear problem (1), accurate to $O(\lambda^4)$, is given by the following formula:

$$p_{+\infty} \equiv p(z = \pi) \approx \frac{P_{RZ}}{4} + \frac{3\lambda^3}{8} C_\infty^3. \qquad (15)$$

This formula describes well the process up to $U_0 \approx 0.3$ ($\lambda \approx 0.1$). After this level is exceeded, significant deviations from the numerical result are observed. This behavior is well understood when inspecting the formula: since $P_{RZ} \leq 1$ and $C_\infty$ does not depend on $U_0$, $p_{+\infty}$ defined by formula (15) grows infinitely as $U_0$ increases, exceeding, already at $U_0 \approx 0.53$ (for $\delta_0 \approx 0$), the maximum value 1/2 allowed by the normalization.

However, the derived formula can be modified to essentially improve the result. This can be done by noting that $p_L$ at small non-zero $\lambda$ is much better approximated by a formula of the form $p_L \approx (P_{RZ}/4)(f/f(t = +\infty)) = P_{RZ} f/(4C_\infty^2)$. This, obviously, corresponds to the choice $A = P_{LZ}/(4C_\infty^2)$ what leads to a formula of significantly better structure:

$$p_{+\infty} \approx \frac{P_{RZ}}{4} + \frac{6\lambda}{C_\infty} \left(\frac{P_{RZ}}{4}\right)^2. \qquad (16)$$

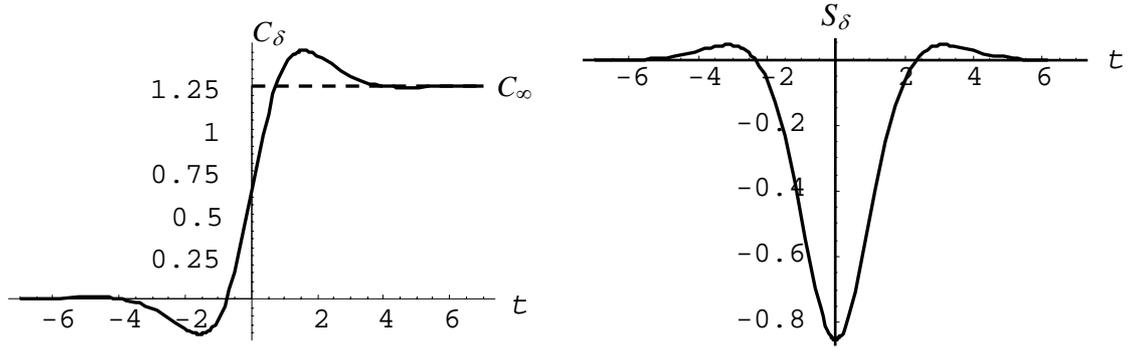

Fig.3. The functions $C_\delta(t)$ and $S_\delta(t)$, $\delta_0 = 0.5$.

Indeed, unlike formula (15), the transition probability $p_{+\infty}$ defined by the given formula remains less than 1/2 under $\lambda \leq 1$, i.e., in that whole range where it makes sense for one to confine himself only to the first term of Picard's series expansion for $u$. However, the obtained formula gives a numerically satisfactory approximation only up to $\lambda \leq 0.15$. Reasons for the latter additional restriction deserve special discussion and we will return to this a little later. But before, we will show that there is a non-trivial way to improve this result even more. Note first that, with accuracy to a constant factor, $F(z) = \lim_{\lambda \to 0}(e^{i\pi/2} a_{2RZ} / \sqrt{\lambda})$. This observation suggests the replacing of the functions $C_\delta$ and $S_\delta$ in (11) by $-\text{Im}(a_{2RZ})/\sqrt{\lambda}$ and $\text{Re}(a_{2RZ})/\sqrt{\lambda}$, respectively, $\cos(\delta(z))$ and $\sin(\delta(z))$ by the corresponding derivatives. As is easily seen, this is nearly equivalent to substitution $C_\infty = \sqrt{P_{RZ}/\lambda}$ in formula (16). As a result, we have

$$p_{+\infty} \approx \frac{P_{RZ}}{4} + 3\left(\lambda \frac{P_{RZ}}{4}\right)^{3/2}. \tag{17}$$

More accurate calculations taking into account the properties of $a_{2RZ}$ show that

$$p_{+\infty} \approx \frac{P_{RZ}}{4} + 3(1+2\lambda)\left(\lambda \frac{P_{RZ}}{4}\right)^{3/2}. \tag{18}$$

The derived formula gives very good approximation up to $\lambda \leq 0.25$ ($U_0 < 0.5$), the relative error is of the order of fractions of a percent, see Fig.4.

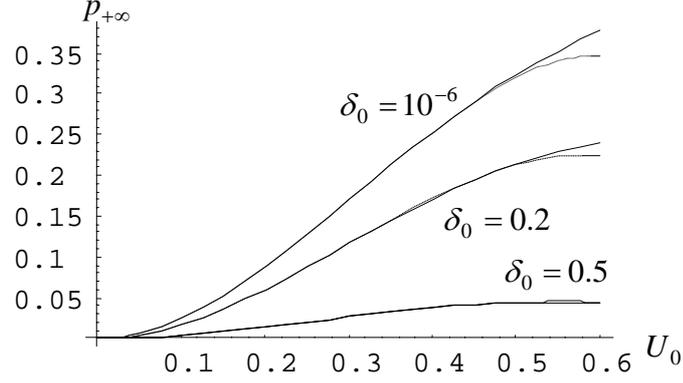

Fig.4. Final probability of the transition to the molecular state as a function of $U_0$: solid line - numerical result, dashed line - Eq. (18).

Let us now discuss the applicability range for the obtained formulas and the origin of the restriction imposed on $\lambda$. The calculations above rest upon the presumption of smallness of Picard's successive approximations for $u$ as compared to the first term of Picard's series. As follows from Eq. (11), the second Picard's term has the form

$$u_1 = -4\lambda \int_0^z K(z,\eta) \left[ (1-3p_L)u_0 - \frac{3}{2}u_0^2 \right] d\eta. \tag{19}$$

As it is immediately seen, whenever at $\lambda \ll 1$ the condition $u_0 \sim \lambda$ [$p_L = O(\lambda)$] is fulfilled, the assumption $u_1 \ll u_0$ is warranted to be the case. Of course, this takes place under $\lambda \leq 0.1$ and, as was mentioned above, it is this fact that defines the applicability range of formulas (15) and (16). The situation, however, is drastically changed already at $\lambda \approx 0.2 \div 0.3$. First, one should no longer consider $\lambda$ as being much less than unity, and, second, what is more important, the linear transition probability $p_L$ is not any longer much less than unity. The latter is already seen from Rosen-Zener formula (5): the final probability of linear transition $p_L(t=+\infty) = P_{RZ}/4$ at $\delta_0 \approx 0$ is about 0.25. Therefore, strictly speaking, as *weakly-nonlinear* cases (when the solution in the zero-order approximation is given by the linear one) in Rosen-Zener model (2) under consideration, one should regard those ones whose dimensionless amplitudes of the field obey the condition $U_0 \leq 0.3$ ($\lambda \leq 0.1$). Note that this conclusion is not *a priori* evident. For instance, one can compare with the Landau-Zener model where the weak nonlinearity limit corresponds to the values of $\lambda < 1$ [12]. Thus, fields with $U_0 > 0.3$ ($\lambda > 0.1$) belong to the intermediate type between the strong-nonlinear and weak nonlinear ones. It is for this reason that modifications (17) and (18) applicable up to $U_0 < 0.5$ ($\lambda \leq 0.25$) are of substantial importance. One might hope that the latter formula will be applicable for a little larger $\lambda$ if $\delta_0 \geq \lambda$, since then, due to the presence of the factor $[\text{sech}(\pi\delta_0)]^2$ in formula (5), $p_L(t = +\infty) \ll 1$. To some extent, this assumption is valid, however, this time the relative error is significantly greater and amounts to an order of several percent already at $U_0 \approx 0.7$. This is because the Rosen-Zener linear solution $p_{2RZ}(t)$ [see (4)] in the vicinity of the point $t = 0$ reaches values which are of the order of unity, irrespective of values of $\delta_0$ (at moderate and large values of $\delta_0$, there is a pronounced maximum). This is not difficult to see by considering, e.g., the linear solution behavior at $U_0 = 1$ when the hypergeometric series in solution (4) are terminated and an elementary solution, $p_{2RZ} = (\text{sech}(t))^2 /(1+4\delta_0^2)$, is obtained. As is seen, at moderate $\delta_0 \approx 0.5$ holds $p_{2RZ}(0) \approx 0.5$. Thus, the general conclusion is that under $U_0 > 0.5 \div 0.6$, one may not confine himself only to the first term of Picard's series for $u$ since the successive terms play an important role. Thereby, the given regime should be viewed as a strongly nonlinear one.

## 3. Strong coupling limit

In the strong coupling limit of high field intensities, $U_0^2 \gg 1$, the nonlinearity is well pronounced. In this case, however, the Volterra equation (8) is of little help, because the successive Picard's approximation terms become larger and larger. Instead, we use the following exact nonlinear differential equation of the third order [22]

$$p_{ttt} - \left(\frac{\delta_{tt}}{\delta_t} + 2\frac{U_t}{U}\right)p_{tt} + \left[\delta_t^2 + 4U^2(1-3p) - \left(\frac{U_t}{U}\right)_t + \frac{U_t}{U}\left(\frac{\delta_{tt}}{\delta_t} + \frac{U_t}{U}\right)\right]p_t + \frac{U^2}{2}\left(\frac{\delta_{tt}}{\delta_t} - \frac{U_t}{U}\right)(1 - 8p + 12p^2) = 0. \quad (20)$$

For the Rosen-Zener model under consideration the frequency detuning is constant, and the equation is well simplified

$$p_{ttt} + 2\tanh(t)p_{tt} + [(4\delta_0^2 + 1) + 4U_0^2 \text{sech}^2(t)(1-3p)]p_t + \frac{U_0^2}{2}\text{sech}^2(t)\tanh(t)(1 - 8p + 12p^2) = 0. \quad (21)$$

To construct an approximate solution to this equation, compare the magnitudes of involved terms keeping in the mind that we suppose $U_0^2 \gg 1$. It is then immediately seen that there are two basic possibilities depending on the magnitude of the detuning, $\delta_0 \ll 1$ and $\delta_0 \gg 1$. This conclusion is also guessed from Fig.1. Indeed, as was already noted above, at small detuning the final conversion probability (i.e., the molecular state probability at $t \to +\infty$) reveals large amplitude oscillatory dependence on the Rabi frequency. In the meanwhile, the probability rapidly decreases as the detuning is increased becoming practically negligible at $\delta_0 \approx 1$. These observations are further confirmed by examining the time evolution of the transition probability (Fig.5). We see that at $\delta_0 \leq 0.5$ strong atom-molecule time-oscillations occur (see the detailed picture in Fig.6), while at larger detuning the oscillations are highly suppressed (Fig.7); they can be neglected already at $\delta_0 \approx 2$.

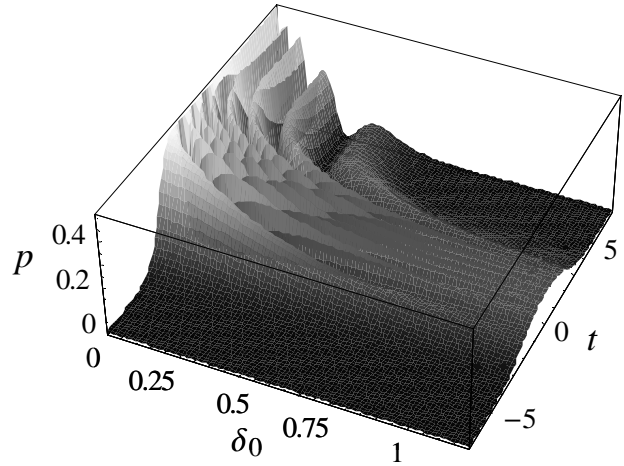

Fig. 5. Molecule formation probability vs. time at different detuning ($U_0 = 10$).

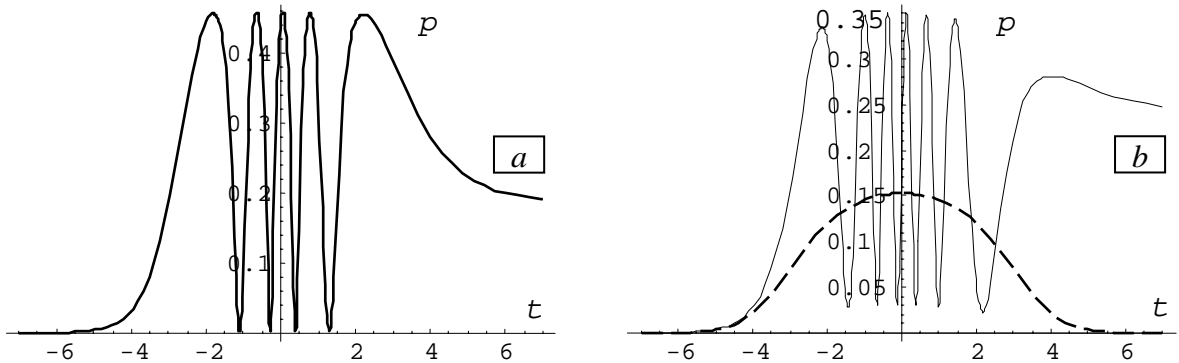

Fig. 6. Molecular state time evolution at small detuning: a) $\delta_0 = 0.05$, b) $\delta_0 = 0.2$ ($U_0 = 10$). Solid line - numerical solution, dashed line - limit solution (28).

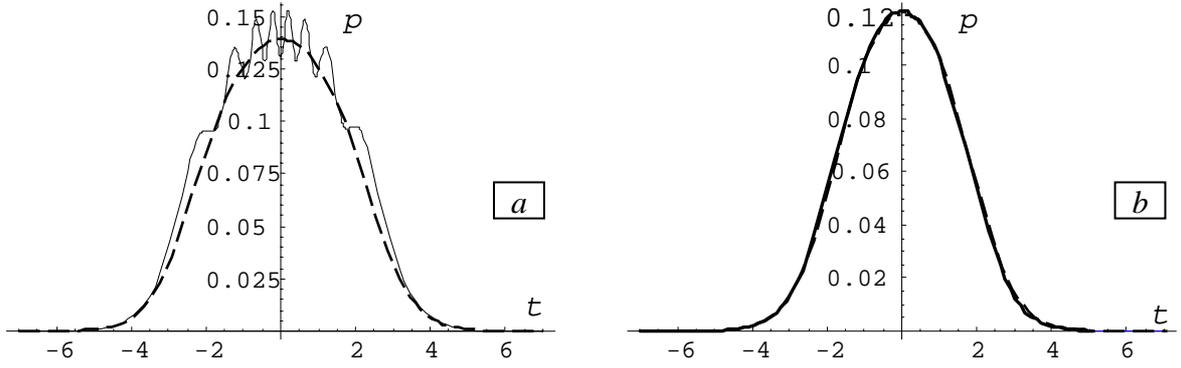

Fig. 7. Molecular state probability vs. time at large detuning: a) $\delta_0 = 1$, b) $\delta_0 = 2$ ($U_0 = 10$). Solid line - numerical solution, dashed line - limit solution (28).

***Large detuning case:*** $\delta_0 \gg 1$.

Since both $U_0$ and $\delta_0$ are large parameters, the leading terms in Eq. (21) are the last two. Keeping only these terms and denoting thus constructed solution by $p_0$ we get the following ***limit*** equation

$$[(4\delta_0^2 + 1)\cosh^2(t) + 4U_0^2(1 - 3p_0)]p_{0t} + \frac{U_0^2}{2}\tanh(t)(1 - 8p_0 + 12p_0^2) = 0. \qquad (22)$$

This equation is solved by a change of the independent variable followed by interchange of the roles of the independent and dependent variables. Indeed, the transformation

$$\frac{d}{dt} = \alpha \tanh(t) \cdot s \frac{d}{ds} \quad \Leftrightarrow \quad \sqrt{4\delta_0^2 + 1}\cosh(t) = C_0 s^{1/\alpha} \qquad (23)$$

changes Eq. (22) to the form

$$s\alpha \frac{dp_0}{ds}[C_0^2 s^{2/\alpha} + 4U_0^2(1 - 3p_0)] + \frac{U_0^2}{2}(1 - 8p_0 + 12p_0^2) = 0. \qquad (24)$$

Choosing now $\alpha = -2$ and $C_0 = U_0$ we arrive at an equation,

$$[1 + 4s(1 - 3p_0)]\frac{dp_0}{ds} - \frac{1}{4}(1 - 8p_0 + 12p_0^2) = 0, \qquad (25)$$

that can be further solved by considering $s$ as a dependent variable because then the equation becomes linear. The result reads

$$\frac{U_0^2}{(4\delta_0^2 + 1)\cosh^2(t)} = \frac{C_1 + p_0(p_0 - 1/2)^2}{9(p_0 - 1/2)^2(p_0 - 1/6)^2}. \qquad (26)$$

For the initial condition $p_0(-\infty) = 0$ considered here holds $C_1 = 0$, hence the equation is considerably simplified reducing to a quadratic equation for $p_0$:

$$\frac{U_0^2}{(4\delta_0^2+1)\cosh^2(t)} = \frac{p_0}{9(p_0-1/6)^2}, \qquad (27)$$

whereby we arrive at the following principal result:

$$p_0(t) = \frac{1}{6} + \frac{4\delta_0^2+1}{18U_0^2}\cosh(t)\left(\cosh(t) - \sqrt{\cosh^2(t) + \frac{6U_0^2}{4\delta_0^2+1}}\right). \qquad (28)$$

This is a highly accurate approximation. For $U_0 > 5$ and $\delta_0 > 2$ the probability calculated by this formula and the numerical result are practically indistinguishable (Fig. 7b). Besides, it allows one to linearize Eq. (21) (by substitution $p = p_0 + u$) thus covering the whole range $\{U_0 \geq 1, \delta_0 \geq 1\}$. Two immediate conclusions follow from this formula. First, since $p_0(t \to +\infty) \to 0$, the final molecular state probability at strong coupling is nearly zero if the detuning is large, i.e., the system posed to a large-detuning Rosen-Zener pulse returns to its initial all-atomic state. Second, $p_0(t)$ is a bell-shaped non-oscillatory function of time and its maximum is achieved at $t=0$:

$$p_0^{\max} = \frac{1}{6} + \frac{4\delta_0^2+1}{18U_0^2}\left(1 - \sqrt{1 + \frac{6U_0^2}{4\delta_0^2+1}}\right). \qquad (29)$$

For $6U_0^2/(4\delta_0^2+1) \gg 1$ this is close to $1/6$. However, importantly, $p_0^{\max}$ is always less than $1/6$. Hence, at large detuning a Rosen-Zener pulse is not able to associate more than one third of atoms ($p_{\text{molecule}} = 1/6$ corresponds to the $1/3$ of atoms). This limitation for the conversion efficiency had been noted to be the case in the adiabatic limit (which is equivalent to the discussed case of high field intensities and large detuning) for other non-crossing models too (see, e.g., [23]). Another variation of the $1/3$ limitation is the observation that for the crossing models in the adiabatic approximation the molecular state probability is always close to $1/6$ at the resonance crossing time-point (see, e.g., [12, 22]).

Finally, it is of interest to compare the above nonlinear behavior under strong coupling and large detuning conditions with the linear Rosen-Zener counterpart. In the linear case, instead of Eq. (22), we have the following linear limit equation (here, the normalization $I_L = 1/4$ is adopted)

$$[(4\delta_0^2+1)\cosh^2(t) + 4U_0^2]p_{0Lt} + \frac{U_0^2}{2}\tanh(t)(1-8p_{0L}) = 0. \qquad (30)$$

The solution to this equation reads

$$p_{0L}(t) = \frac{1}{8}\left(1 - 1/\sqrt{1 + \frac{4U_0^2}{4\delta_0^2 + 1}\mathrm{sech}^2(t)}\right). \tag{31}$$

This formula demonstrates the same qualitative features as the nonlinear solution, Eq. (28); i.e., in the linear case again a return to the initial state is observed if the applied Rosen-Zener pulse is of a large detuning, and there is a maximal possible transition probability achieved at $t = 0$. This time, this probability is $1/8$ (i.e., $1/2$ for normalization $I_L = 1$).

*Small detuning case:* $\delta_0 \ll 1$.

To treat this regime we first rewrite Eq. (21) in the following factorized form

$$\left(\frac{d}{dt} + \tanh(t)\right)\left[p_{tt} + \tanh(t)p_t - \frac{U_0^2 \mathrm{sech}^2(t)}{2}(1 - 8p + 12p^2)\right] + 4\delta_0^2 p_t = 0. \tag{32}$$

The speculations now to proceed are as follows. Though the detuning is supposed to be small, one cannot completely neglect the term $4\delta_0^2 p_t$. Indeed, putting $\delta_0 = 0$ results in a monotonically increasing solution:

$$p = \frac{1}{2}\tanh^2\left(\frac{U_0 z}{\sqrt{2}}\right), \tag{33}$$

where

$$z = \frac{\pi}{2} + 2\arctan(\tanh(t/2)), \quad z \in [0, \pi]. \tag{34}$$

However, the numerical simulations reveal that for any nonzero small $\delta_0$ the solution is oscillatory (this is well seen from Figs. 5 and 6). Hence, in a sense, the exact resonance case $\delta_0 = 0$ is degenerate. This degeneracy can be resolved by introducing a small perturbation when constructing the initial approximation. Intuitively, in order to get an approximation that is as close to the real solution as it is possible, one should try to introduce a perturbation as small as possible. On the other hand, one should choose a form of this perturbation that allows construction of an analytic solution. From this point of view, the form of Eq. (32) suggests to introduce the perturbation inside the square brackets since then the truncated equation, the equation that remains after disregarding the term $4\delta_0^2 p_t$, is immediately integrated once. One may further try to choose a specific form of the perturbation that allows complete integration of the reduced equation. The listed requirements are all achieved by introducing in square brackets in Eq. (32) a trial term of the form $A \cdot \mathrm{sech}^2(t)$ with some constant $A$ (depending, in general, on $\delta_0$ and $U_0$). We suppose that $A$ is small (say, of the order of $\delta_0^2$ as $\delta_0$ goes to zero). The value of this constant is then defined by requiring the resultant approximation to be as close to the exact solution as possible with the chosen form of the introduced perturbation.

To proceed with the outlined approach, we rewrite Eq. (32) in the following equivalent form:

$$\left(\frac{d}{dt} + \tanh(t)\right)\left[p_{tt} + \tanh(t)p_t - \frac{U_0^2 \text{sech}^2(t)}{2}\left(1 - 8p + 12p^2 - \frac{2A}{U_0^2}\right) - A \cdot \text{sech}^2(t)\right] + 4\delta_0^2 p_t = 0. \quad (35)$$

Now, supposing $A \ll 1$, we neglect the last two terms of this equation and integrate the remaining equation once. Taking into account the initial conditions applied here, we arrive at the following second order equation

$$p_{tt} + \tanh(t)p_t - \frac{U_0^2 \text{sech}^2(t)}{2}\left(1 - 8p + 12p^2 - \frac{2A}{U_0^2}\right) = 0, \quad (36)$$

which is easily turned into the one with constant parameters by the change of the independent variable given by Eq. (34):

$$p_{zz} - \frac{U_0^2}{2}\left(1 - 8p + 12p^2 - \frac{2A}{U_0^2}\right) = 0. \quad (37)$$

Multiplying this equation by $p_z$ and integrating once we obtain an equation that is immediately identified as an equation for an effective Rabi problem for some field parameters – for an effective field amplitude and an effective detuning (see, e.g., [8]). Correspondingly, the zero-order approximation is written in terms of the Jacobi elliptic sine function [20]:

$$p_0 = p_1 \text{sn}^2[\sqrt{p_2}U_0 z; m], \quad (38)$$

where parameters $p_1, p_2$ and $m$ are defined as

$$p_{1,2} = \frac{1}{2} \mp \sqrt{\frac{A}{2U_0^2}}, \quad m = \frac{p_1}{p_2}. \quad (39)$$

Comparing this solution with the exact resonant solution we first note that the solution given by Eq. (33) is also written in terms of Jacobi sn-function if one takes $m = 1$ ($\tanh(z) = \text{sn}[z;1]$). Furthermore we note that Eq. (38) is reduced to Eq. (33) for $A = 0$. These observations clearly suggest that the performed procedure, the introduction of an $A$-term, is equivalent to changing the parameters of the resonant solution (33) written in the Jacobi sn-function form. Hence, the approach we applied can be viewed as a modification of the well-known method of strained parameters [24].

The obtained solution (38) presents an oscillatory function the behavior of which displays all the qualitative features of the exact solution. Moreover, a few numerical simulations shortly reveal that for any small enough $\delta_0$ one may always find such a value of the parameter $A$ for which this solution is practically indistinguishable from the numerical solution. To derive an analytic expression for this value of $A$ examine the neglected terms with $p$ defined by this solution:

$$R = \left(\frac{d}{dt} + \tanh(t)\right)\left[-A \cdot \mathrm{sech}^2(t)\right] + 4\delta_0^2 p_{0t} = \mathrm{sech}^2(t)\left[A \cdot \tanh(t) + \frac{4\delta_0^2 p_{0t}}{\mathrm{sech}^2(t)}\right]. \quad (40)$$

Here, the idea is to choose the parameter $A$ so that this remnant becomes as small as possible. Strictly speaking, one should look for a value of $A$ for which the ***influence*** of the neglected terms is minimal. To address the latter question mathematically strongly, one should examine the behavior of the next approximation term constructed by means of using $p_0$ of Eq. (38) as zero-order approximation. However, it is difficult to proceed in this way because the analytic expression for the next approximation term is not known. For this reason, we look for indirect criteria. A possibility opens up when examining the behavior of function $4\delta_0^2 p_{0t}/\mathrm{sech}^2(t)$. This is a step-wise function that exponentially slowly decreases from a relatively large value $\sim \delta_0^2 U_0^2$ at $t = -\infty$, then sharply goes to zero at some negative time-point and remains negligible at a large vicinity of the point $t = 0$ where the field intensity is the highest. Noting now that a rather similar qualitative behavior is displayed by the term $A \cdot \tanh(t)$, we see that the remnant $R$ will by essentially suppressed for a large time-interval, covering the effective interaction region, i.e., the vicinity of the point $t = 0$, if we require the term embraced in the square brackets in Eq. (40) to vanish at the beginning of the interaction, i.e. at $t \to -\infty$. Then, since

$$4\delta_0^2 p_{0t}\big|_{t \to -\infty} \sim 8\delta_0^2 p_1 p_2 U_0^2 \mathrm{sech}^2(t), \quad (41)$$

we immediately get

$$A = \frac{2\delta_0^2 U_0^2}{1 + 4\delta_0^2}. \quad (42)$$

This is already a good approximation showing the order of the parameter $A$: $A \sim \delta_0^2 U_0^2$. Indeed, the comparison with the numerical solution shows that the approximate solution (38) with this value of $A$ well describes the process for many oscillations (see Fig. 8a).

Nevertheless, it is seen that the deviation from the exact solution slowly increases during the time and eventually becomes rather notable at the end of the interaction process. Fortunately, the result can be essentially improved by trying a perturbation with two fitting parameters, namely a perturbation of the form $(A+Bp)\text{sech}^2(t)$. Since then the parameters $p_1$ and $p_2$ of Eq. (38) are changed independently [compare with Eq. (39)], it is understood that this is more elaborate realization of the strained parameters method. Interestingly, it turns out that for high field intensities the result is effectively equivalent to the single-parameter $A$-perturbation approach with a slightly modified value of $A$ as compared with that of Eq. (42):

$$A = \frac{\sqrt{2}\,\delta_0^2 U_0^2}{1+4\delta_0^2}. \tag{43}$$

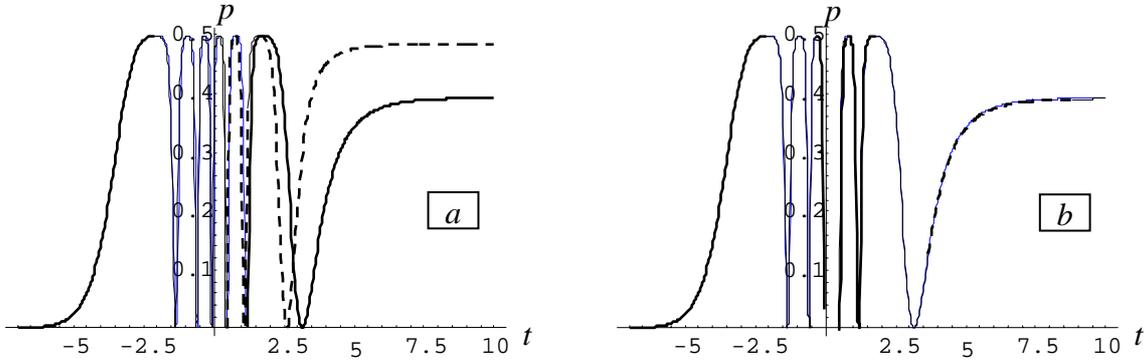

Fig. 8. Comparison of the approximation (38) (dashed line) with numerical result (solid line) for $\delta_0 = 0.001$, $U_0 = 23.5$: a) $A$ is given by Eq. (42) and b) $A$ is given by Eq. (43).

The parameters $p_1$ and $p_2$ are finally given by simple formulas:

$$p_{1,2} \approx \frac{1}{2} \mp \frac{\delta_0}{\sqrt{\sqrt{2}}}. \tag{44}$$

This is a really good approximation. The Jacobi sine solution (38) with these parameters produces graphs practically indistinguishable from the numerical solution as far as $\delta_0$ is small enough and $U_0 \gg 1$ (Fig. 8b). If needed, one may further improve the results by linearization of the problem using this solution as an initial approximation.

Thus, we have seen that at small detuning the Rosen-Zener pulse causes large amplitude oscillations during the time evolution of the coupled atom-molecule ensemble described by the Jacobi sn-function. According to the properties of this function, the shape of the oscillations is defined by the parameter $m = p_1/p_2$. Hence, we conclude from Eq. (44) that at small detuning the time-shape of the atom-molecule oscillations is in the first approximation defined by the detuning only. On the other hand, the number of the oscillations is mainly defined by the value of $U_0 z = \int_{-\infty}^{+\infty} U_0 \text{sech}(t)\, dt$, i.e., by the pulse area.

It is interesting to analyze the way the above approximation for small detuning was constructed from a different point of view. We have seen that the constant $A$, which determines both the qualitative and quantitative properties of the solution, was eventually calculated by examining the behavior of the system at the beginning of the interaction. This observation leads to a notable speculation. Indeed, it seems rather unexpected that the vicinity of the point $t = -\infty$ where the amplitude of the field is exponentially small plays such an important role, a much more important role than that of the vicinity of point $t = 0$, where the field amplitude is maximal. This clearly indicates that the time point $t = -\infty$ actually presents a hidden singularity. The origin and the nature of this singularity are understood by rewriting Eq. (35) for the variable $z$ (constant-amplitude form of the equation):

$$\left(\frac{d}{dz} - \frac{\delta_{zz}}{\delta_z}\right)\left[p_{zz} - \frac{U_0^2}{2}(1 - 8p + 12p^2)\right] + \delta_z^2 p_z = 0 , \qquad (45)$$

where $z = \int_{-\infty}^{t} \text{sech}(t)\, dt$ and the effective detuning $\delta_z$ is given as

$$\delta_z(z) = 2\delta_0 / \sin(z) \qquad (46)$$

so that

$$\frac{\delta_{zz}}{\delta_z} = -\frac{\cos(z)}{\sin(z)} . \qquad (47)$$

It is then immediately seen from the last relation that the point $z = 0$ corresponding to $t = -\infty$ is indeed singular because the operator $\delta_{zz}/\delta_z$ diverges at this point. Notably, this divergence does not depend on the parameter $\delta_0$ which is the only characteristic of the detuning $\delta(t)$. The divergence is of course caused by the transformation from $t$ to $z$, hence, exclusively by the form of the time-evolution of the field amplitude, $U(t)$, more precisely, by the speed of inclusion of the field. Naturally, this singularity can be viewed as an effective resonance-touching (but not crossing) because the divergence of the operator $\delta_{zz}/\delta_z$ at the crossing point is the main characteristic of the (constant-amplitude) crossing models (e.g., for the Landau-Zener case we have $\delta_{tt}/\delta_t = 1/t \to \infty$ at $t = 0$). Strictly speaking, there is another singular point, $z = \pi$, however, this corresponds to $t = +\infty$, where the interaction process ends. Since the interaction exponentially vanishes as approaching to this point and there is no further time for this point to display its influence, the role of this point is in practice negligible. Note finally that the left-hand side of Eq. (26) which describes the other evolution regime corresponding to the large detuning case can be written as $\approx 4U_0^2 / \delta_z^2$. Since this term vanishes at $z = 0$, thus leading to a zero integration constant $C$, we conclude that in this regime too the behavior of the system is essentially determined by the mentioned hidden singularity.

## 4. Summary

We have examined, in the limits of two-modes' Gross-Pitaevskii mean field approach, the molecule formation process in a Bose-Einstein condensate under the conditions of the non-crossing Rosen-Zener model for which the detuning of the field is constant and the pulse amplitude is varied according to the hyperbolic secant law.

We have first studied the weak coupling limit for this field configuration. Using an exact nonlinear Volterra integral equation, we have shown that in this limit the solution to the problem is written in terms of the solution to an auxiliary linear Rosen-Zener problem. We have derived a simple expression for the final transition probability. We have found that for the Rosen-Zener model the strict limit of weak nonlinearity corresponds to smaller field intensities than for other known models such as the Landau-Zener and Nikitin-exponential ones. We have shown that this is because of the inherit properties of the particular hyperbolic secant pulse shape under consideration.

Further, we have treated the strong coupling limit of high field intensities when the nonlinearity is most pronounced in the molecule formation process. We have shown that here there are two different regimes of the time evolution of the coupled atom-molecule system corresponding to large and small detuning of the associating field. In the first case the behavior of the system is almost non-oscillatory while in the second case large amplitude coherent oscillations in the population dynamics are observed.

Discussing the large detuning regime, we have shown that the conversion process is effectively described by a limit first-order nonlinear equation for the molecular state probability. Using the exact solution to this equation, we have shown that in this regime the molecular fraction qualitatively follows the field amplitude time-variation, i.e., the probability of the molecular state first monotonically increases, reaches a maximum at the time point when the field intensity is maximal, and then decreases as the field amplitude decreases. Eventually, the system returns to the initial all-atomic state. The maximal possible molecular fraction is found to be $1/6$, i.e., in this regime a Rosen Zener pulse is capable to capture no more than the third of the initial atomic population (this is an argument why a resonance-crossing is needed for molecule production efficiency). In accordance with this prediction, the JILA experiments [25] have shown a maximum molecular conversion of about 16%.

Furthermore, discussing the small detuning limit, we have shown that in this time the system is well described by a second order nonlinear equation that is shown to be the equation for an effective Rabi problem with changed parameters. We have derived accurate approximations for the parameters of the corresponding Rabi-solution written in terms of the Jacobi elliptic sine function. We have seen that the number of the oscillations, as in the linear case, is mainly defined by the pulse area. In the meantime, we have shown that the oscillation shape is mostly defined by the field detuning; the influence of the field intensity here presents a small correction of higher order.

Finally, we have indicated an inherent singularity of the Rosen-Zener model, a hidden singularity that stands for many of the qualitative and quantitative properties of the model. This singularity, which is shown to be due to the time-variation law of the field amplitude at the beginning of the interaction, can be viewed as an effective resonance-touching.

**Acknowledgments**

This work was supported by the International Science and Technology Center Grant A-1241, the Academy of Finland Project No. 115682 and INTAS Young Scientist Fellowship Ref. No. 06-1000014-6484. A.I. thanks the Laboratoire de Physique Moléculaire et des Collisions, Université Paul Verlaine - Metz (France), R.S. and A.I. thank the Department of Physics, University of Turku (Finland) for kind hospitality.